\documentclass[
 reprint,twocolumn,
 amsmath,amssymb,
 aps,nofootinbib,superscriptaddress,
]{revtex4-2}

\usepackage{amsthm}
\usepackage{graphicx}
\usepackage{dcolumn}
\usepackage{bm}
\usepackage{color}
\usepackage{braket}

\newcommand{\bx}{\ensuremath{\boldsymbol{x}}}

\newcommand{\bw}{\ensuremath{\boldsymbol{w}}}
\newcommand{\br}{\ensuremath{\boldsymbol{r}}}
\newcommand{\bu}{\ensuremath{\boldsymbol{u}}}
\newcommand{\bp}{\ensuremath{\boldsymbol{p}}}
\newcommand{\bP}{\ensuremath{\boldsymbol{P}}}
\newcommand{\bQ}{\ensuremath{\boldsymbol{Q}}}
\newcommand{\bX}{\ensuremath{\boldsymbol{X}}}
\newcommand{\bSigma}{\ensuremath{\boldsymbol{\Sigma}}}

\newcommand{\cD}{\ensuremath{\mathcal{D}}}
\newcommand{\cE}{\ensuremath{\mathcal{E}}}
\newcommand{\cU}{\ensuremath{\mathcal{U}}}
\newcommand{\cT}{\ensuremath{\mathcal{T}}}
\newcommand{\cR}{\ensuremath{\mathcal{R}}}

\newcommand{\uth}{^{\rm th}} 

\newcommand{\Ctot}{C_{\rm tot}}
\newcommand{\Ctottiv}{C_{{\rm tot}}^{\rm TIV}}
\newcommand{\Ctottv}{C_{{\rm tot}}^{\rm TV}}

\newcommand{\Ctotdtiv}{C_{{\rm tot},d}^{\rm TIV}}
\newcommand{\Ctotdtv}{C_{{\rm tot},d}^{\rm TV}}

\newcommand{\nrmse}{{\rm NRMSE}}
\newcommand{\limT}{\lim_{T\rightarrow\infty}}

\DeclareMathOperator{\tr}{\textup{Tr}}

\usepackage{hyperref}
\hypersetup{
    colorlinks = true,
    linkcolor = {blue},
    citecolor = {blue},
    urlcolor = {blue},
}

\begin{document}

\preprint{APS/123-QED}

\title{Quantum Noise-Induced Reservoir Computing}

\author{Tomoyuki Kubota$^*$}
\email{kubota@ai.u-tokyo.ac.jp}
\affiliation{Next Generation Artificial Intelligence Research Center (AI Center), Graduate School of Information Science and Technology, The University of Tokyo, Japan}

\author{Yudai Suzuki$^*$}
\email{yudai.suzuki.sh@gmail.com}
\affiliation{Department of Mechanical Engineering, Keio University, Japan}

\author{Shumpei Kobayashi}
 \affiliation{Department of Creative Informatics, The University of Tokyo, Japan}
 
\author{Quoc Hoan Tran}
 \affiliation{Next Generation Artificial Intelligence Research Center (AI Center), Graduate School of Information Science and Technology, The University of Tokyo, Japan}
 
\author{Naoki Yamamoto}
 \affiliation{Quantum Computing Center, Keio University, Japan}
 \affiliation{Department of Applied Physics and Physico-Informatics, Keio University, Japan}

\author{Kohei Nakajima}
 \affiliation{Next Generation Artificial Intelligence Research Center (AI Center), Graduate School of Information Science and Technology, The University of Tokyo, Japan}
 \affiliation{Department of Creative Informatics, The University of Tokyo, Japan}
 \affiliation{Department of Mechano-Informatics, The University of Tokyo, Japan}
\date{\today}

\begin{abstract}
Quantum computing has been moving from a theoretical phase to practical one, presenting daunting challenges in implementing physical qubits, which are subjected to noises from the surrounding environment.
These quantum noises are ubiquitous in quantum devices and generate adverse effects in the quantum computational model, leading to extensive research on their correction and mitigation techniques.
But do these quantum noises always provide disadvantages?
We tackle this issue by proposing a framework called quantum noise-induced reservoir computing and show that some abstract quantum noise models can induce useful information processing capabilities for temporal input data.
We demonstrate this ability in several typical benchmarks
and investigate the information processing capacity to clarify the framework's processing mechanism and memory profile.
We verified our perspective by implementing the framework in a number of IBM quantum processors and obtained similar characteristic memory profiles with model analyses.
As a surprising result, information processing capacity increased with quantum devices' higher noise levels and error rates.
Our study opens up a novel path for diverting useful information from quantum computer noises into a more sophisticated information processor.
\end{abstract}

\maketitle

\def\thefootnote{*}\footnotetext{These authors share first authorship.}\def\thefootnote{\arabic{footnote}}

Quantum computing has been viewed as a futuristic technology, shifting from a fantastical perspective to practical applications.
In this paradigm, reducing computational errors in quantum hardware due to the noisy surrounding environment has become the biggest obstacle~\cite{georgescu:2020:25years}.
These noise effects are the main hindrance in building large-scale future quantum systems.
However, they motivate the realization of the quantum advantage under the limited implementation of noisy intermediate-scale quantum (NISQ) computers~\cite{preskill:2018:NISQ,cerezo:2021:var}.
In the literature, noise is ubiquitous, not limited to quantum systems, and has been extensively studied in dynamical systems.
Rather than simply obscuring certain deterministic dynamics, some of the noise-induced effects can stabilize the dynamics~\cite{matsumoto1983noise}, admit chaos~\cite{crutchfield1982fluctuations}, synchronize~\cite{maritan1994chaos}, 
or exhibit important information to characterize biological and physical systems~\cite{horsthemke:1983:transition,ridolfi:2011:phenomena,jhawar:2020:noise-schooling}. An intriguing research question arises: Can the quantum noise that causes physical qubits to lose their quantum mechanical properties induce positive effects for information processing?

\begin{figure*}
    \centering
    \includegraphics[width=16cm]{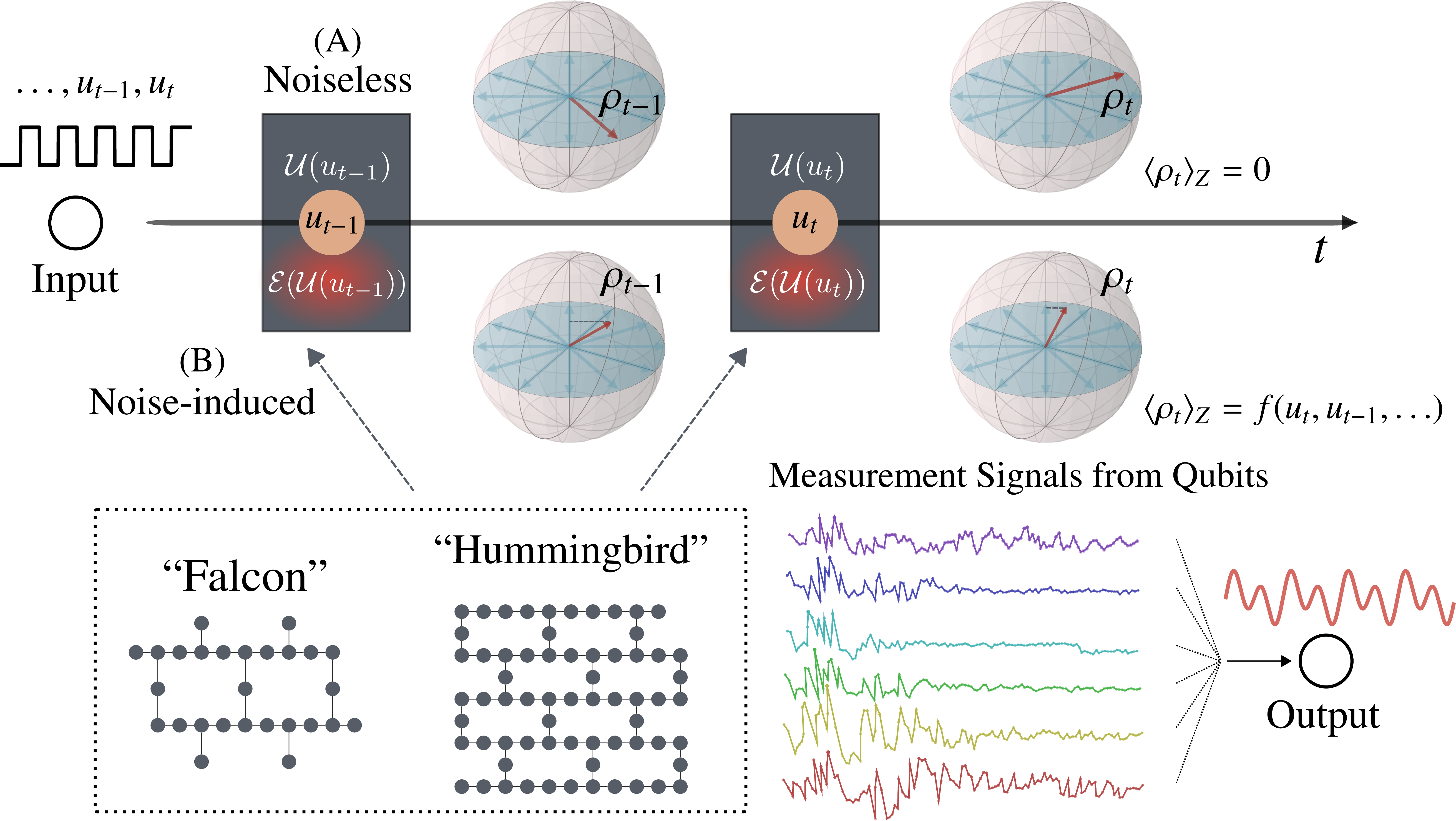}
    \caption{
        Overview of how to leverage quantum noises as computational resources in the abstract model and real quantum processors.
        In a noiseless evolution with input-dependent unitary $\cU(u_t)$, measurement results at each time step may not carry information about the input trajectories. For example, if $\cU$ is designed such as if the Bloch vector only rotates in the equatorial plane and measurements are performed in the Pauli Z basis, then measurement results are always zero, regardless of inputs [panel (A)]. In a noisy evolution [panel (B)], the quantum noise channel $\cE$ may induce the meaningful processing where measurement results become the function of input trajectories, leading to the formation of a quantum noise-induced reservoir (QNR). These measurement results are expected to carry transformed information of input history in a high-dimensional state space and can be used to emulate the target function of input trajectories via a simple linear combination.
        From this perspective, the natural noise in specific quantum processors can be used to construct the QNR.
    }
    \label{fig:overview}
\end{figure*}

Several studies have tried to exploit quantum noise as a useful resource in specific applications: 
to induce universal quantum computation by engineering the local dissipation~\cite{verstraete:2009:noise:disp},
to prepare quantum thermal states with high fidelities by introducing and optimizing parameterized noise models~\cite{foldager:2021:noise-assisted},
and to enhance the robustness of quantum classifiers by placing depolarization noise in the quantum circuits~\cite{du:2021:noise-adversaries}.  
Even with these positive usages, there have been no studies on which type of mechanism induces an expected computation from quantum noise and how to quantify the power of such mechanism as an information processing component.

We propose a general model of quantum noise-induced temporal information processing via the reservoir computing (RC) framework and define this as a quantum noise-induced reservoir (QNR).
Here, reservoir refers to an input-driven dynamical system that maps sequential inputs into a high-dimensional state space where the previous inputs information can be retained~\cite{jaeger:2001:echo,maass2002real,jaeger2004harnessing,lukoeviius:2009:reservoir,nakajima:2021:RC}.
As a result, we can use a simple linear combination of these states to approximate target functions of input trajectories.
This framework can be extended to  physical reservoir 
computing, which exploits physical dynamics as a computational resource~\cite{nakajima2020physical}.
Due to the high degree of freedom, quantum systems such as 
disordered quantum spins~\cite{fujii2017harnessing,nakajima2019boosting,tran:2020:higherorder,tran:2021:temporal}, 
fermions or bosonic networks~\cite{ghosh:2019:quantum,ghosh:2019:neuromorphic,ghosh:2020:reconstruct,khan:2021:qrc:boson},
harmonic oscillators~\cite{nokkala:2020:gaussian,gerasimos:2021:prx:measurement},
and photonic quantum memristors~\cite{spagnolo:2022:natphotonic}
can present good reservoirs.
Particularly in dissipative quantum systems, some proof-of-principle experiments have been demonstrated on IBM quantum processors~\cite{chen2020temporal,suzuki2021natural}.
The dissipative property can originate from the natural noise in a specific circuit, forming a natural quantum reservoir (QR)~\cite{suzuki2021natural}.

Our QNR stems from an input-dependent noiseless quantum circuit model where the sequence of measurement results does not contain information on the input trajectories. 
We can give this circuit the ability of information processing by applying noise to it, such as the amplitude damping noise or unintended interactions between nearby physical qubits.
We then demonstrate its usage in a benchmark of time-series regression and a practical indirect sensing problem. 
To evaluate the power of such mechanism, we quantify the information processing ability of QNR via a novel and powerful tool called \textit{temporal information processing capacity} (TIPC)~\cite{kubota2021unifying}, which is used for the first time on quantum systems.
Here, TIPC measures the capacity of the input-driven dynamics to reconstruct polynomial functions of input and the reservoir's internal state history.
TIPC provides insights into the temporal processing mechanism and tells us what noise models can truly induce a required computational capability for a temporal processing task.
Furthermore, given a function of input trajectories, we can use TIPC to evaluate the required properties to learn this function, such as what combinations of past inputs are processed. 
Finally, we implement the natural QR~\cite{suzuki2021natural} in various IBM quantum processors and anticipate that the natural QR belongs to our QNR model.
Surprisingly, their TIPC profiles show that quantum processors with more noisy effects, such as higher error rates, tend  to exhibit better performance as a QR in temporal processing tasks.

\section*{Quantum Noise-induced Reservoir} 
To build a QNR, we start from the QR framework based on the quantum dissipative system~\cite{fujii2017harnessing,chen:2019:dissipative}.
Consider the input $\{u_t\}$ and target sequence $\{y_t\}$ where $t$ denotes the time step and $y_t$ is a function of a finite input history $u_t, \ldots, u_{t-t_0}$. 
The goal of a temporal learning task is to emulate the relationship between $\{u_t\}$ and $\{y_t\}$.
Let $\rho_t$ be the quantum state of the system at time $t$; 
the Markovian quantum dissipative system is represented as
\begin{align}
\rho_t = \cT_{u_t}(\rho_{t-1}),
\end{align} 
where $\cT_{u_t}$ is an input-dependent completely positive trace preserving (CPTP) map.
The measurement signals are obtained via $N$ observables $O_1, O_2, \ldots, O_N$ measuring on $\rho_t$ as $x_{i,t}={\rm Tr}(\rho_t O_i)$. 
Here, $\bx_t=[x_{1,t}\cdots x_{N,t}]^\top$ is called the reservoir state at time step $t$. 
The prediction output is described via a readout family of linear combination $\hat{y}_{t} = \hat{\bw}^\top \bx_t$,
where the weight $\hat{\bm w}$ is learned by the least squares method on a classical computer, meaning $\hat{\bm w} = \arg \min_{\bm w}\left[ \sum_t (y_t-{\bm w}^\top {\bx}_t)^2 \right]$.

As shown in Ref.~\cite{suzuki2021natural}, we can easily give an example of $\cT_{u_t}$ and the observables such that the reservoir states do not contain the information of input.
Assuming the QR system consists of $N=2n$ qubits, we choose $\cT_{u_t}$ as
\begin{align}~\label{eqn:unitary:map}
    \rho_{t} = \cT_{u_t}(\rho_{t-1}) = \cU(u_t)\rho_{t-1}\cU(u_t)^\dag,
\end{align}
where the unitary operator $\cU(u_t)$ is defined as
\begin{align} \label{eq:unitary}
\cU(u_t) = U_{0,1}(u_t)\otimes U_{2,3}(u_t)\otimes\cdots\otimes U_{N-2,N-1}(u_t).
\end{align}
Here, given the input-scaling coefficient $s$, we consider
\begin{align}
    U_{i,j}(u_t) = \textup{CX}_{i,j} \textup{RZ}_j(s u_t) \textup{CX}_{i,j} \textup{RX}_i(s u_t) \textup{RX}_j(s u_t), \label{eq:unitary_2qubit}
\end{align}
where $\textup{CX}_{i,j}$ represents the CNOT gate with the control qubit $i$ and the target qubit $j$, and $ \textup{RZ}_i(\theta)$ ($ \textup{RX}_i(\theta)$) represents the \textup{RZ} (\textup{RX}) gate that rotates a single-qubit state labeled $i$ through angle $\theta$ on the Z (X) axis. 
If the initial state is set as $\rho_0 = \ket{+_n}\bra{+_n} = H^{\otimes n}\ket{{\bm 0}}\bra{{\bm 0}} H^{\otimes n}$ with $\ket{{\bm 0}} = \ket{0}^{\otimes n}$, and the Pauli $Z$ basis on each qubit is chosen as the observables, the reservoir states ${\bx}_t = {\bm 0}$ for all $t$, regardless of inputs. 
We chose the observables $Z_i = I\otimes \cdots \otimes I \otimes Z \otimes I \otimes \cdots \otimes I$, where the single qubit Pauli Z operator is placed at the $i$th index.

We construct noise channels to induce a useful computation of input sequence, which is intuitively illustrated in Fig.~\ref{fig:overview}.
In the noiseless situation, the Bloch vectors of the quantum system are only processed on the equatorial plane, making the Pauli Z measurement result in an input-independent value.
In the presence of a quantum noise channel $\cE$, the reservoir states may carry information of the input trajectories under the Pauli Z measurement. If $\cE$ causes the Bloch vector to deviate from the equatorial plane, the measurement result may become an input-dependent value.
We modify the unitary dynamics in Eq.~\eqref{eqn:unitary:map} to include noise channels 
\begin{align}
    \rho_{t} = \mathcal{E}_{\mathrm {deco}}(\mathcal{N_{\mathrm{unitary} }}\left[\cU(u_t)\right]\rho_{t-1} \mathcal{N_{\mathrm{unitary}}}\left[\cU(u_t)\right]^\dag).
    \label{eq:sim_dynamics}
\end{align}
Here, $\mathcal{E}_{\mathrm {deco}}$ is a combination of multiple decoherence noise channels of different types, which includes amplitude damping, phase damping, uniform depolarization, bit-flip and phase-flip. 
We also consider the unitary noise $\mathcal{N}_{\mathrm{unitary}}$, 
where $\mathcal{N}_{\mathrm{unitary}}[\cU(t)]$ is a unitary operator modified from $\cU(t)$,
including the modification to mimic unintended entanglement effects between nearby qubits (see Methods). 

\begin{figure}
    \centering
    \includegraphics[width=8.5cm]{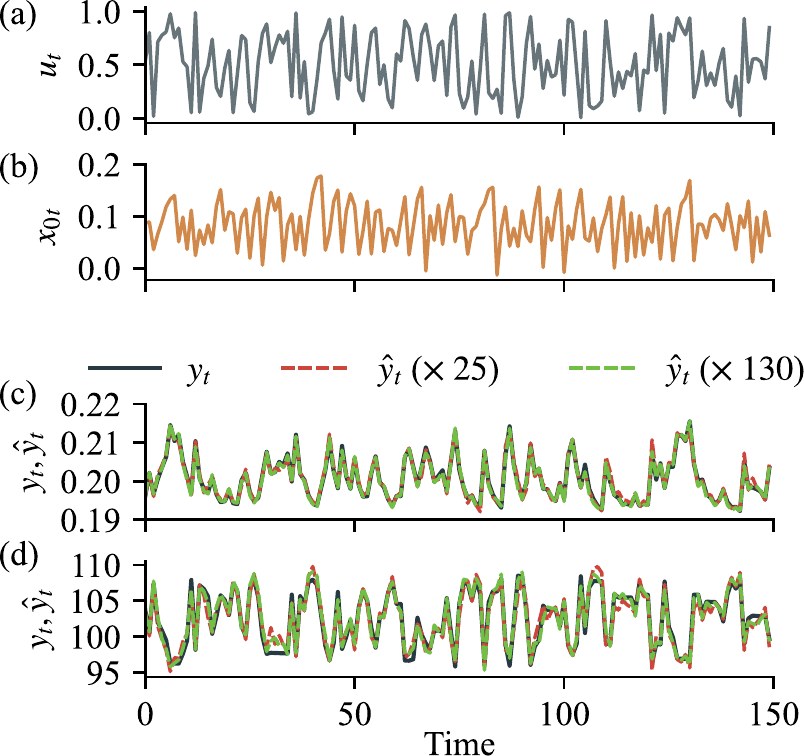}
    \caption{
        The benchmark tasks solved by spatial multiplexing of 4-qubit QNRs. (a) Input sequence. (b) Typical reservoir state time-series. 
        (c) The target of NARMA2 task (black) and outputs (red, $4~{\rm qubits}\times25$; green, $4~{\rm qubits}\times130$). (d) The target of the PAM length emulation task (black) and outputs (red, $4~{\rm qubits}\times25$; green, $4~{\rm qubits}\times130$). 
    }
    \label{fig:benchmark}
\end{figure}

We perform two temporal learning tasks: the second-order nonlinear autoregressive moving average (NARMA2) benchmark~\cite{atiya2000new} and the pneumatic artificial muscle (PAM) length emulation tasks with the data set provided in Ref.~\cite{akashi2020input}.
The targets are functions of input history, which require memories and nonlinearity to emulate. 
We build the 4-qubit QNR with the input-scaling $s=\pi$ and the 10 types of quantum noises (see Methods).
The combination of these noises leads to total of $1024$ QNR instances.
We use the spatial multiplexing technique~\cite{nakajima2019boosting}, where the reservoir states in different QNRs are combined to learn the target.
The performance is evaluated via the normalized root mean square error 
\begin{align}
\textup{NRMSE} = \dfrac{1}{\sigma(y)}\sqrt{\dfrac{1}{N_{\textup{eval}}} \sum_{t=1}^{N_{\textup{eval}}} (y_t - \hat{y}_t)^2 }.
\end{align}
Here, $\sigma^2(y)$ is the variance of the target sequence and $\hat{y}_t$ is the prediction at time step $t$ in $N_{\textup{eval}}$ time steps.
Figure~\ref{fig:benchmark} shows that the spatial multiplexing of $130$ ($25$) QNRs with a total of 520 (100) computational nodes can emulate the target sequences with high precision [NARMA2, $\nrmse=0.11~(0.21)$; PAM length, $\nrmse=0.21~(0.30)$]. 
Their performances for NARMA2 are equivalent to those of the conventional echo state network~\cite{jaeger:2001:echo} (ESN) in classical RC with 110 (50) computational nodes (see Methods). For the PAM task, the QNR slightly outperformed the ESN with less than 520 nodes ($\nrmse>0.22$),
motivating us to investigate the information processing components that make the QNR perform better than ESN on some specific tasks.

\section*{Temporal Information Processing Capacity}

\begin{figure*}
    \centering
    \includegraphics[width=17cm]{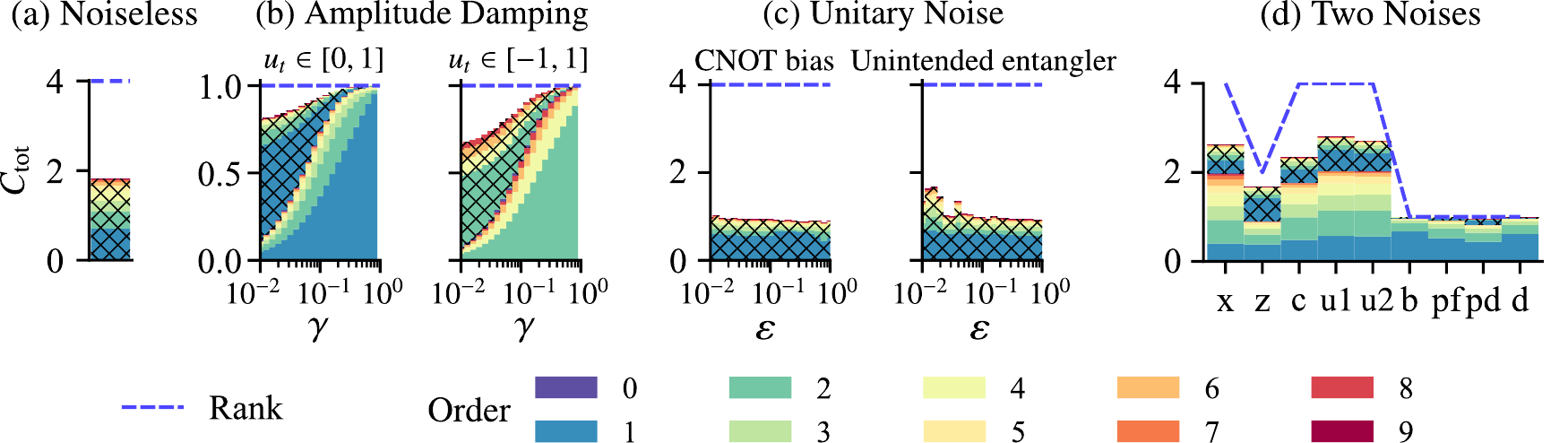}
    \caption{
        TIPC decomposition of $4$-qubit QNR models in Eq.~\eqref{eq:sim_dynamics} with: 
        (a) the ideal circuit without quantum noise;
        (b) amplitude damping (left: $u_t\in[0,1]$; right: $u_t\in[-1,1]$) noise varying by damping rate $\gamma$; 
        (c) unitary noises (left: CNOT bias for over-rotation; right: unintended entangling between nearby qubits to mimic unintended entanglement effects) varying by the pertubated rate $\varepsilon$;
        and (d) the composition of amplitude damping with another noise [Pauli X (x), Pauli Z (z), CNOT bias (c), one-hop (u1) and two-hop (u2) unintended entangler, bit-flip (b), phase-flip (pf), phase damping (pd), or depolarization (d) noise ($\varepsilon=\gamma=0.1$)]. The dotted blue lines represent the ranks of the reservoir states in each reservoir setting.
        The hatched bars represent the time-variant components in TIPC, while the other parts represent the time-invariant components. In (a),(c), and (d), the input $u_t$ is uniformly distributed in $[0, 1]$.
    }
    \label{fig:TIPC_qubit:sim}
\end{figure*}

\begin{figure*}
    \centering
    \includegraphics[width=17cm]{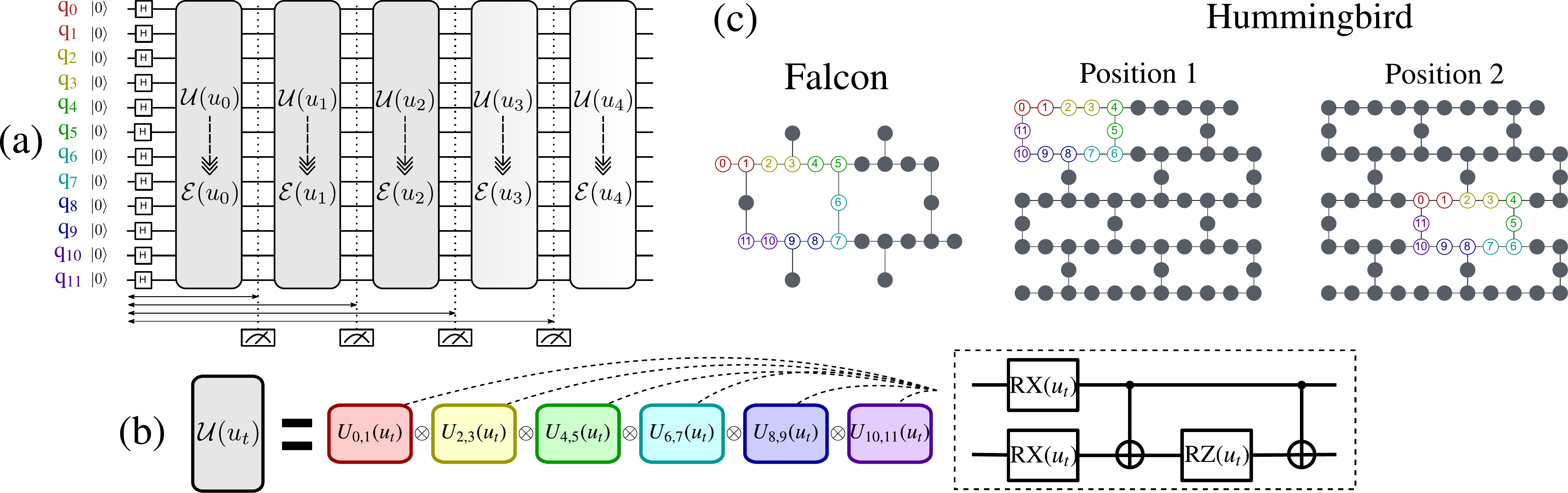}
    \caption{
    The realization of the QNR model on IBM superconducting quantum processors. 
    (a) The dynamics is constructed by an input-independent unitary evolution $\cU(u_t)$ with RX, RZ, and CNOT gates (b), where the Pauli Z measurement results are constant regardless of inputs, showing meaningless computation in the ideal situation. Quantum noises existing in the quantum processors can drive the input-driven dynamics, enabling the processing of the input sequence.
    (c) The position of QNR's qubits (the nodes with colored labels) in each real quantum processor. Each subsystem applied by the unitary $U_{i,i+1}(u_t)$ in (b) composes the qubits of the same color.  
    }
    \label{fig:real-machine}
\end{figure*}

In temporal learning tasks, we assume that target output $y_t$  is described as a function of a finite input history 
\begin{eqnarray}
    y_{t} &=& f(u_{t},\ldots,u_{t-t_0}), 
\end{eqnarray}
where $f(\xi_0, \ldots, \xi_{t_0})$ is a time-invariant representation for variables $\xi_0, \ldots, \xi_{t_0}$.
In general, the QNR's states $\bx_t=(x_{1,t}\cdots x_{N,t})$ can be described not only by input history but also by time $t$ as follows: 
\begin{eqnarray}\label{eqn:t-dependent}
    x_{i,t} &=& g_i(t, u_{t},\ldots,u_{0};\rho_{0}). 
\end{eqnarray}

\begin{figure*}
    \centering
    \includegraphics[width=17cm]{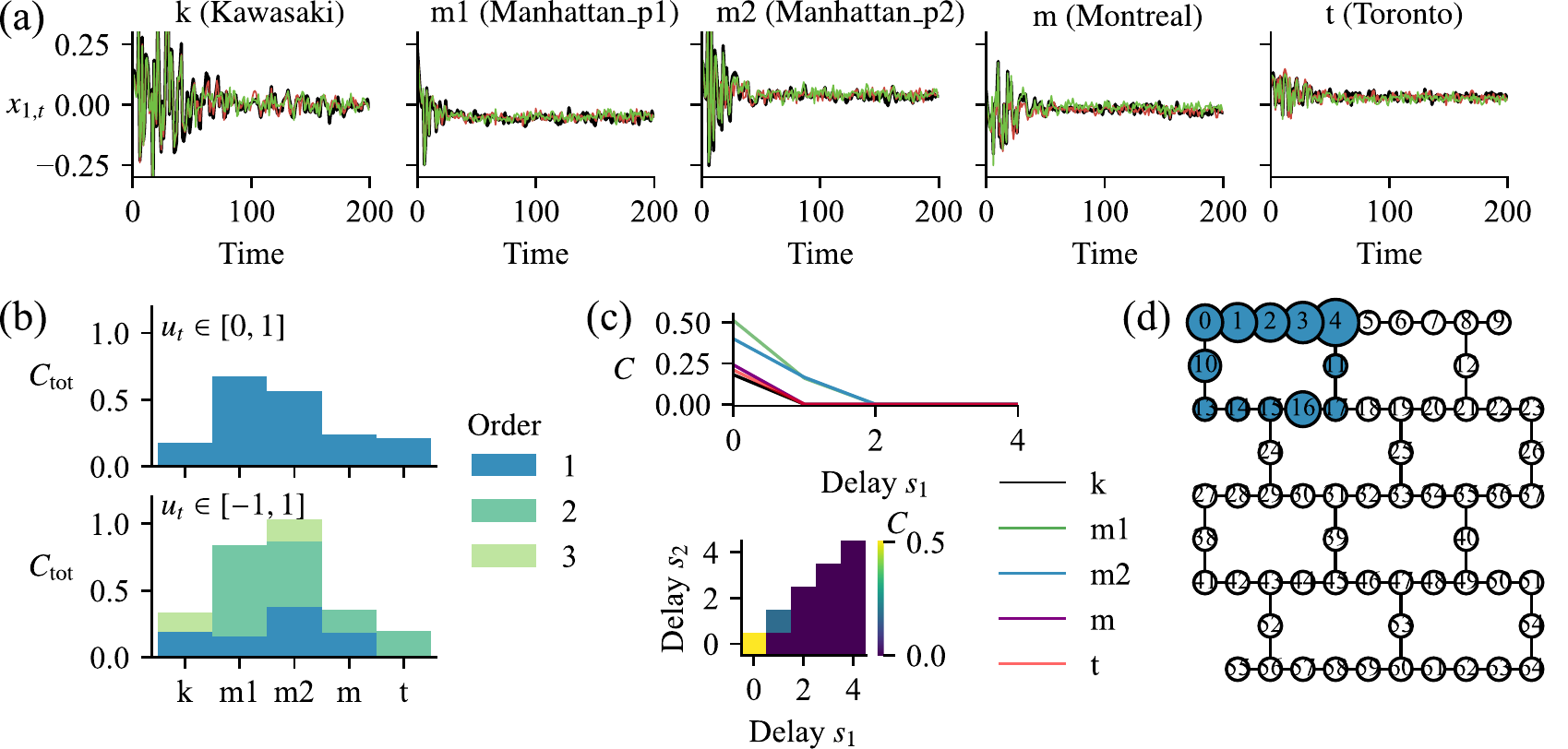}
    \caption{
        TIPC decomposition of natural QRs~\cite{suzuki2021natural}, which can be considered as QNRs implemented in real quantum processors. 
        (a) The representative reservoir states time-series $x_{1,t}$ of five quantum machines [Kawasaki (k), Manhattan\_p1 (m1), Manhattan\_p2 (m2), Montreal (m), and Toronto (t)] for three trials (black, red, and green). 
        (b) The averaged TIPC decomposition of machines (upper, $u_t\in[0,1]$; lower, $u_t\in[-1,1]$). 
        (c) Time-invariant first-order capacities from five machines (upper, $u_t\in[0,1]$) and the second-order time-invariant capacities from Manhattan\_p1 (lower, $u_t\in[-1,1]$), respectively. 
        (d) The total of time-invariant first-order capacities (represented by the size of nodes) contributed in each qubit of Manhattan\_p1. 
    }
    \label{fig:TIPC_nQR}
\end{figure*}

We use TIPC~\cite{kubota2021unifying} 
to evaluate the capacity to reconstruct polynomial functions of input and reservoir state history
via characterizing the time-invariant and time-variant factors in the orthonormal basis expansion of $g_i$.
Using singular value decomposition (SVD), we first transform $N$-dimensional state $\bx_{t}$ into $r$-normalized linearly independent state $\hat{\bx}_{t}$, where $r$ denotes the rank of the covariance matrix of states (see Methods). 
The $r$-dimensional orthonormal state is expanded by orthonormal bases $z_{j,t}=z_j(t,u_t,\ldots,u_{t_0})$ that depend on time and input history.
These basic functions compose a complete orthonormal system as follows: 
\begin{align}~\label{eqn:decompose}
    \hat{\bx}_{t} = \sum_j {\bm\gamma}_j z_{j,t} = \sum_j {\bm\gamma}_j z_{j}(t,u_{t},\ldots,u_{t_0}).
\end{align}

The TIPC for the $i$th term is described by $C_i = ||{\bm\gamma}_i||^2$. If the bases construct a complete system, the orthonormality yields the completeness property that the total capacity $\Ctot=\sum_i C_i$ is equivalent to $r$. 
Therefore, the TIPC can provide a comprehensive description of temporal information processing. 
Furthermore, the $d$th-order TIPC decomposition for time-invariant terms $\Ctotdtiv$ and time-variant terms $\Ctotdtv$ are defined by 
\begin{eqnarray}
    \Ctotdtiv &=& \sum_{\{j|N_j=d,M_j=0\}} C_j, \\
    \Ctotdtv &=& \sum_{\{j|N_j=d,M_j>0\}} C_j,
\end{eqnarray}
where $N_j$ and $M_j$ represent the orders of input and internal state in Eq.~\eqref{eqn:decompose}, respectively (see Methods).
We emphasize that $\Ctotdtiv$ enables the temporal processing capability in emulating the function of input trajectories, while $\Ctotdtv$ hinders this capability due to the unreproducible processing.
We further define the total capacities for time-invariant and time-variant terms as $\Ctottiv = \sum_d\Ctotdtiv$ and $\Ctottv=\sum_d\Ctotdtv$.

Figure~\ref{fig:TIPC_qubit:sim} depicts the TIPC profiles for the simulation of the $4$-qubit QNR in the noiseless situation 
[Fig.~\ref{fig:TIPC_qubit:sim}(a)], the amplitude damping noise [Fig.~\ref{fig:TIPC_qubit:sim}(b)], the unitary noise defined by CNOT over-rotation (CNOT bias) and unintended entangling of nearby qubits [Fig.~\ref{fig:TIPC_qubit:sim}(c)], and the compositions of amplitude damping noise with another type of noise [Fig.~\ref{fig:TIPC_qubit:sim}(d)].
In some situations, the total capacities $\Ctot$ did not saturate the ranks $r$ since we discarded sufficient small $C_i$ values that were less than a predefined threshold  (see Methods).
In the circuits with noiseless or unitary noise channels, such as CNOT bias and unintended entangler, the TIPC is only constructed of time-variant components.
These remind us that no information of the input sequence is carried in these settings.
In the models with a one-noise model of single-qubit unitary, phase-flip, bit-flip, phase damping, or depolarization, $\Ctot=r=0$.
In contrast, the amplitude damping noise clearly induces time-invariant components in the TIPC profiles, both setting it alone~[Fig.~\ref{fig:TIPC_qubit:sim}(b)] and in combination with other noises~[Fig.~\ref{fig:TIPC_qubit:sim}(d)].
Time-invariant components become dominant as the amplitude damping rate $\gamma$ increases, 
implying that higher noise error rates tend to induce better QNR performance.

\section*{QNR in real quantum processors}

The hardware-specific noise in real quantum processors can be used to construct the QNR model, and their TIPC can provide insights into the information processing capability of the real system. 
We refer to the natural QR system introduced in Ref.~\cite{suzuki2021natural} for the implementation. We anticipate that this natural QR belongs to our QNR model with the dynamics of
\begin{eqnarray}
    \rho_{t} = \mathcal{E}_{\rm device}(\cU(u_t)\rho_{t-1} \cU(u_t)^\dag), \label{eq:cptpmap_device}
\end{eqnarray}
where $\cU(u_t)$ is the input-dependent unitary defined in Eqs.~\eqref{eqn:unitary:map} and \eqref{eq:unitary} with input-scaling $s=2$, and $\mathcal{E}_{\rm device}(\cdot)$ denotes the unknown hardware-specific noise channel. 
We utilized two families of IBM quantum processors called Hummingbird and Falcon, which have 65 and 27 qubits, respectively. 
We adopted two configurations of 12-qubit QNRs in different positions of an \textit{ibmq\_manhattan} device (denoted as the Manhattan\_p1 and Manhattan\_p2) for the Hummingbird type and the \textit{ibm\_kawasaki}, \textit{ibmq\_montreal} and \textit{ibmq\_toronto} devices (denoted as the Kawasaki, Montreal and Toronto, respectively) for the Falcon type 
[Fig.~\ref{fig:real-machine}(c)].
We used the Qiskit framework~\cite{Qiskit} to perform the experiments.

Figure~\ref{fig:TIPC_nQR}(a) shows representative reservoir states time-series obtained from three trials of the five machines under the same input sequence. 
We computed the TIPCs of these QNRs and confirmed that
the total capacity $\Ctot$ and time-invariant capacity $\Ctottiv$ depend on the type of machine. 
As shown in Fig.~\ref{fig:TIPC_nQR}(b), only time-invariant capacities appear, where the Hummingbird-type device has larger capacities than the Falcon-type devices.
The first-order capacities contribute as major parts for asymmetric input $u_t\in[0,1]$. 
The second-order capacities dominate with symmetric input $u_t\in[-1,1]$, but first-order capacities still appear in several machines.
Since there is no first-order capacity in the symmetric input case with the QNR constructed in the simulation [Fig.~\ref{fig:TIPC_qubit:sim}(b)], these capacities in real machines may come from other than amplitude damping noises.
In Fig.~\ref{fig:TIPC_nQR}(c), the upper panel shows the first-order capacity as a function of delay step $s_1$. This suggests the short-term memory effect where the reservoir state mainly reflects recent inputs $u_{t-s_1}~(s_1=0,1,2)$. 
The lower panel illustrates the time-invariant second-order capacities in Manhattan\_p1 for the orthonormalized term of $u_{t-s_1}u_{t-s_2}$, where $u_t^2$ dominates. 
In Fig.~\ref{fig:TIPC_nQR}(d),  the colored node's size depicts the total of time-invariant first-order capacities computed with each qubit in Manhattan\_p1 (see Supplemental Materials for other results). 

\begin{figure}
    \centering
    \includegraphics[width=8.5cm]{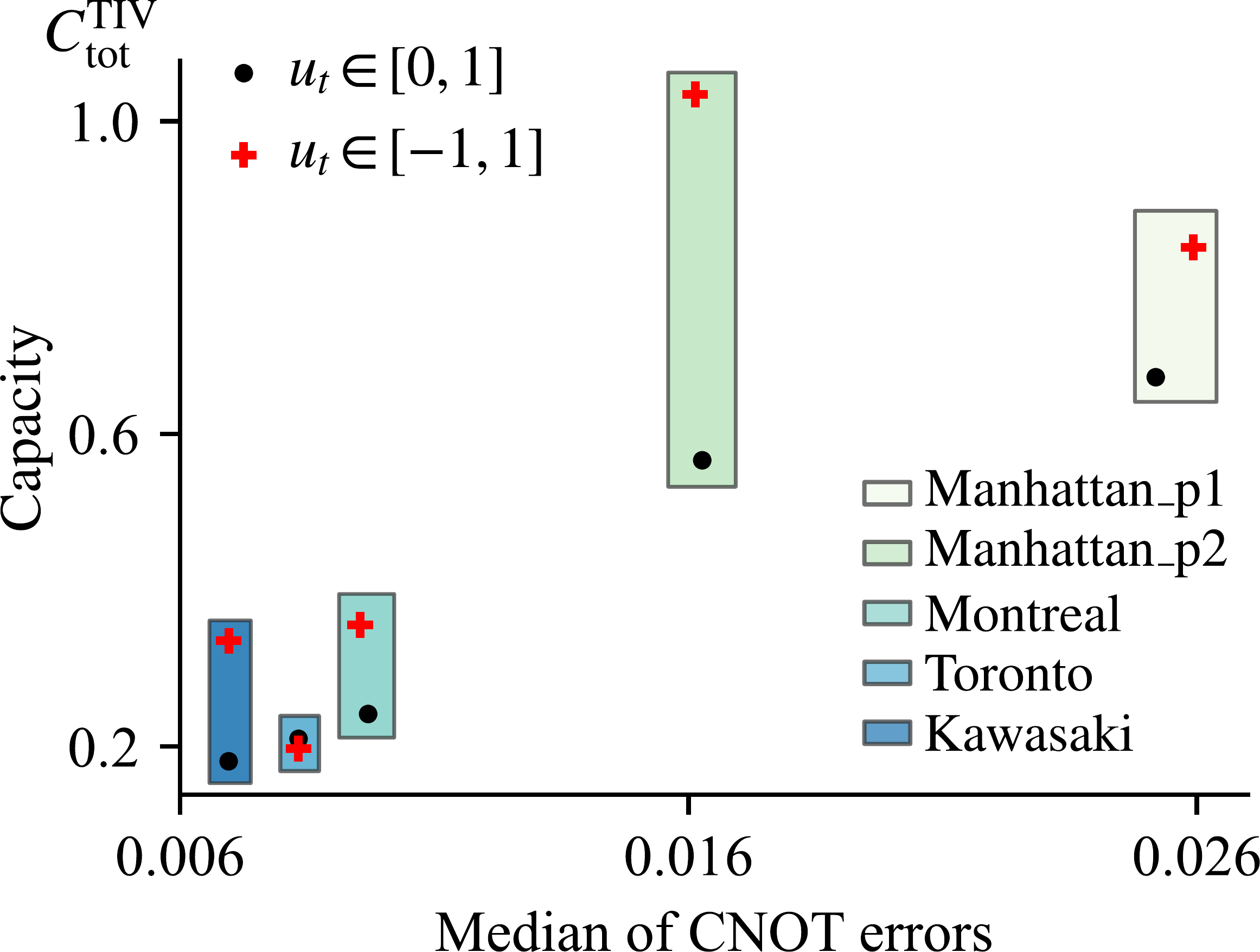}
    \caption{
    Relationship between CNOT errors of IBM quantum machines and the total time-invariant capacity $\Ctottiv$ averaging over 10 trials of QNR models implemented in these machines. 
    The rectangles summarize two points of each machine ($u_t\in[0,1]$, black dot; $u_t\in[-1,1]$, red plus).
    Full results are presented in Table~\ref{tab:hardware_parameters}.
    }
    \label{fig:CNOT-all-machine}
\end{figure}

We investigated the relationship between CNOT error rates for each quantum machine and the total time-invariant capacity $\Ctottiv$ (Fig.~\ref{fig:CNOT-all-machine}).
We again stress that the time-invariant TIPCs are the main ingredients for the useful information processing of the input sequence.
Surprisingly, the performance of Manhattan devices were the worst in the view of CNOT errors but exhibited the highest capacities in the view of QNR processing.
These measures showed a positive correlation, implying an intriguing way to utilize unavoided and ubiquitous noises in NISQ devices as useful computational resources.
Figure~\ref{fig:CNOT-all-machine} and Fig.~\ref{fig:TIPC_qubit:sim}(b) exhibit
the consistency in the simulation and real implementation that noisy quantum machines with higher error rates are likely to exhibit better temporal processing abilities in the QNR model.
We conjecture that the time-invariant TIPCs obtained from real quantum computers may be induced by unavoidable dissipation in quantum processors. 

\section*{Possible Extensions}
We conclude that quantum noise can induce temporal processing abilities of input data in the reservoir computing framework.
We characterized these abilities via the recently proposed TIPC tool, which is used here first for quantum systems.
We discuss several outcomes of our proposals and possible extensions for future research.

First, we demonstrated the effective usage of several quantum noise models to construct the QNR model. It remained  open to design an optimal QNR structure for a given task, such as the ansatz of input-dependent unitary operators in noiseless condition and the combination of different noise models. We can consider TIPC as a powerful tool for characterizing the profile of an input\text{--}output map for a given specific task. This profile can be used as a good indicator to design the QNR model, including what type of memory and nonlinearity and how many QNRs we need to construct.

Second, highlighting the positive usage of noise in the NISQ era in this study also paves the novel way to understanding the mechanism of real quantum noise from an information-processing perspective.
Without precise knowledge of noise models, we can infer the detailed properties of quantum noise, which is often seen as an unavoidable black box in complicated quantum circuits.
For instance, the profile of noise-induced computation via TIPC can be helpful in a reverse engineering process to detect the noise-dominated connection or noisy qubits in the quantum processors.
Furthermore, in a temporal processing task using quantum computers, TIPC can be used in post-processing to mitigate the noise's effects on the computation results, such as detecting and removing the unexpected time-variant signals.

Third, the quantum noise itself does not carry any information of the input data, but noise is introduced to process meaningful information processing in quantum systems.
This perspective reminds of the phenomenon that  quantum signals cannot be ampliﬁed without the introduction of noise in the ampliﬁed states. 
Interestingly, this opens the opportunity to define QNR models with input-driven dynamics by artificially adding quantum noises or placing noisy circuits in real quantum systems.
We can utilize quantum noise to control the information processing in the quantum system, which can be identified as a quantum advantage to be distinct from quantum speedups.
This belongs to the general idea of designing a target quantum system by actively engineering the environment that interacts with the system~\cite{mankei:2010:noise-cancel}.

In a broader view, the TIPC method does not limit the analysis of the noise-induced computation in quantum processors but can be applied to general input-driven systems to characterize the underlying temporal classical input\text{--}output map.
This can be further processed to investigate the dynamical regime or phase transition.
For quantum systems, especially in quantum devices, one can consider the quantum map associated with quantum input and output trajectories. 
This leads to a challenging topic in developing the theory of TIPC to incorporate with quantum maps and knowing how the noise of a quantum device affects this type of temporal quantum information processing.

\providecommand{\noopsort}[1]{}\providecommand{\singleletter}[1]{#1}%
%

\section*{Methods}
\subsection*{Dissipative Quantum System As Reservoir}
We explain the QR framework based on the quantum dissipative system~\cite{chen:2019:dissipative}.
Considering the quantum state $\rho_t$ of the system at time step $t$,
the quantum dissipative system is represented as 
\begin{align}
\rho_t = \cT_{u_t}(\rho_{t-1}),
\end{align} 
where $\cT_{u_t}$ is an input-dependent completely positive trace preserving (CPTP) map.

The main philosophy here is we can use the dissipative quantum dynamics for computation without precise tuning or optimization of its internal parameter.
Therefore, the proper design for $\cT_{u_t}$ via the quantum circuit is the crucial factor for enabling the temporal processing ability of the QR.
This design requires three properties~\cite{maass2002real,chen:2019:dissipative}: the echo state property (ESP)~\cite{jaeger2002tutorial, manjunath2013echo, yildiz2012re}, or the convergence property~\cite{pavlov:2005:conv}; the fading memory~\cite{boyd1985fading}; and the separation property.
Here, the ESP ensures the reproducibility that the reservoir states $\bx_t$ are independent of its initial condition for a sufficiently long $t$. The fading memory property implies that if a small perturbation is added to the input sequence $u_t, \ldots, u_{t-t_0}$ in recent times, then the deviated reservoir state  should be close to the original state $\bx_t$ under a magnitude of the perturbation.
Finally, the separation property implies that the reservoir states can distinguish any two different input sequences. 

\subsection*{Benchmarks}
We use the second-order nonlinear autoregressive moving average (NARMA2) benchmark~\cite{atiya2000new}, which is commonly used for evaluating the
computational capability of temporal processing with time dependency and nonlinear transformation of the inputs.
The target $y_t$ of the NARMA2 is described by 
\begin{eqnarray}
    y_t &=& 0.4y_{t-1} + 0.4y_{t-1}y_{t-2} + 0.6(0.3u_t)^3 + 0.1,
\end{eqnarray}
where $u_t$ is the uniform random input in the range of $[0,1]$.
In addition, we present the pneumatic artificial muscle (PAM) length emulation tasks with the open data set provided in Ref.~\cite{akashi2020input}.
Here, PAM is a practical soft actuator controlled by air pressure resulting in complex dynamics with high dimensionality and nonlinearity.
The length of PAM is conventionally measured by an infrared sensor, but its integration limits the softness of the device. It was demonstrated that an echo state network (ESN) can predict PAM's length with the same accuracy as infrared sensors given pressure values as inputs~\cite{sakurai2020emulatingAS}.

We evaluated the performance with the same uniform random input sequence in [0, 1] of 49,998 time steps for two tasks. We discarded the first 9,998 steps as washout of the transient dynamics and used the next $2\times 10^4$ steps for training and the last  $N_{\textup{eval}}=2\times 10^4$ steps for evaluating the NRMSE.
Comparing with the TIPC profile of benchmark tasks such as NARMA2 and PAM, the TIPC profile in simulations (Fig.~\ref{fig:TIPC_qubit:sim})  and real machines (Fig.~\ref{fig:TIPC_nQR}) can clarify the hidden factors that govern the performance in solving temporal learning tasks (see Supplemental Materials).

\subsection*{Echo State Network}
Echo State Network (ESN) is an artificial recurrent neural network model in RC.
Consider ESN with $N$ computational nodes, and let the $i$th reservoir state and input at time step $t$ be $x_{i,t}$ and $u_t$, respectively.
The state equation of the ESN is described by 
\begin{eqnarray}
    x_{i,t+1} = \tanh\left(\sum_{j=1}^N \rho w_{ij}x_{j,t} + \iota w_{{\rm in},i}u_{t+1} \right),
    \label{eq:esn}\nonumber
\end{eqnarray}
where $\rho$ and $\iota(=0.1)$ are the parameters to control the spectral radius of internal weight connections and the input scaling, respectively.
Here, $w_{{\rm in},i}$ denotes the input weight for the $i$th node, and $w_{ij}$ is the internal weight connecting from the $i$th to $j$th node. 
We set the connection probabilities of the internal and input weights to $0.5$ and $0.1$, respectively, and generate both types of weight using the uniform random number in the range of $[0,1]$. 
Note that we normalize the internal weight matrix $[w_{ij}]$, dividing it by its largest absolute eigenvalue. 

We select $\rho=0.6$ for NARMA2 and $\rho=0.3$ for PAM length emulation as the best tuned parameter in each task.
For each $N(=5,10,\ldots,520)$, we compute NRMSEs using 10 different configurations and compare averaged NRMSEs with that of the QNR. 

\subsection*{Temporal Information Processing Capacity}
To compute the Temporal Information Processing Capacity (TIPC), we adopt the Volterra\text{--}Wiener\text{--}Korenberg series~\cite{korenberg1988identifying} as the orthonormal polynomial expansion composed of input history and the reservoir's state history. 
Let the $N$-dimensional state and input be ${\bx}_t=[x_{1,t}\cdots x_{N,t}]^\top\in\mathbb{R}^N$ and $u_t\in\mathbb{R}$, respectively, and the state is a function of past state time-series and input history---meaning ${\bx}_t = {\bm f}(u_{t-1},u_{t-2},\ldots,x_{1,t-1},x_{1,t-2},\ldots,x_{N,t-1},x_{N,t-2},\ldots)$. 
We consider the states matrix $\bX = [\bx_0\ldots \bx_{T-1}]^\top \in \mathbb{R}^{T\times N}$
with $r$ $(1 \leq r \leq \min \{T, N\})$ is the matrix rank of $\bX^\top\bX$.
We obtain the normalized, linearly independent state $\hat{\bx}_t\in\mathbb{R}^r$ via the compact SVD of $\bX$ as ${\bX}={\bP\bSigma\bm Q}^\top$ (${\bP}\in\mathbb{R}^{T\times r}$, ${\bSigma}\in\mathbb{R}^{r\times r}$, ${\bm Q}\in\mathbb{R}^{N\times r}$), where ${\bP}=[\hat{\bx}_0 \cdots \hat{\bx}_{T-1}]^\top$. 
Here, $\bP$ and $\bQ$ are real orthogonal matrices, and ${\bSigma}$ is the square diagonal matrix with non-negative real entries.
The state $\hat{\bx}_t$ is expanded as
\begin{eqnarray}
    \hat{\bx}_t &=& 
    \sum_{i=1}^\infty {\bm c}_i
    u_{t-1}^{n_1^{(i)}}u_{t-2}^{n_2^{(i)}}\cdots 
    \hat{x}_{1,t-1}^{m_{1,1}^{(i)}}\hat{x}_{1,t-2}^{m_{1,2}^{(i)}}\cdots 
    \hat{x}_{N,t-1}^{m_{N,1}^{(i)}} \hat{x}_{N,t-2}^{m_{N,2}^{(i)}}\cdots, \nonumber
\end{eqnarray}
where ${\bm c}_i\in\mathbb{R}^r$ is the coefficient vector, and 
$z_{i,t}$ (in Eq.~\eqref{eqn:decompose})  denotes the $i\uth$ basis, as long as there is a one-to-one correspondence between $z_{i,t}$ and $u_{t-1}^{n_1^{(i)}}\cdots \hat{x}_{1,t-1}^{m_{1,1}^{(i)}}
\cdots \hat{x}_{N,t-1}^{m_{N,1}^{(i)}}\cdots$. 
Here, $N_j=\sum_{t} n_{t}^{(j)}$ and $M_j=\sum_{k=1}^r\sum_{t} m_{k,t}^{(j)}$ are the orders of inputs and reservoir internal states in this representation, respectively.

Using the Gram\text{--}Schmidt orthogonalization, we can obtain the coefficient vectors ${\bm\gamma}_i\in\mathbb{R}^r$ of orthonormalized bases $\xi_{i,t}$ as
\begin{eqnarray}
    \hat{\bx}_t = \sum_{i=1}^\infty {\bm\gamma}_i \xi_{i,t},~~ 
    C({\bX},{\bm\xi}^{(i)}) = ||{\bm\gamma}_i||^2, \label{eq:tipc}
\end{eqnarray}
where ${\bm\xi}^{(i)}=[\xi_{i,1}\cdots\xi_{i,T}]^\top$ 
and $||{\bm\xi}^{(i)}||=1$.

We explain that the numerical error of TIPC caused by time length follows the $\chi^2$ distribution with $r$ degrees of freedom.
If we denote ${\bP}=[{\bp}_1 \cdots {\bp}_r]$, where ${\bp}_j\in\mathbb{R}^T$ and $||{\bp}_j||=1$, 
Eq.~\eqref{eq:tipc} can be rewritten as 
\begin{eqnarray}
    C({\bX},{\bm\xi}^{(i)}) &=& \sum_{j=1}^r ||{\bp}_j^\top\cdot{\bm\xi}^{(i)}||^2 \nonumber\\
    &=& \sum_{j=1}^r \left( \frac{1}{T}\sum_{t=1}^T (\sqrt{T}p_{j,t}) (\sqrt{T}\xi_t^{(i)}) \right)^2. \nonumber\label{eq:tipc_innerproduct}
\end{eqnarray}
To investigate the numerical error of TIPC caused by a finite length of time-series, we shuffle ${\bm\xi}^{(i)}$ in time direction to make it an i.i.d. random variable $\tilde{\bm\xi}^{(i)}$. 
Here, $\sqrt{T}{\bp}_j$ and $\sqrt{T}\tilde{\bm\xi}^{(i)}$ have zero means 
$\limT\frac{1}{T}\sum_{t=1}^T \sqrt{T}p_{j,t}
=\limT\frac{1}{T}\sum_{t=1}^T \sqrt{T}\tilde{\xi}_t^{(i)}=0$ 
and unit variances 
$\limT\frac{1}{T}||{\sqrt{T}\bp}_j||^2
=\limT\frac{1}{T}||\sqrt{T}\tilde{\bm\xi}^{(i)}||^2=1$. According to their means, variances, and independence, 
\begin{eqnarray}
    \limT\frac{1}{T}\sum_{t=1}^T (\sqrt{T}p_{j,t})^2 (\sqrt{T}\tilde{\xi}_{t}^{(i)})^2=1. \nonumber
\end{eqnarray}
Therefore, the central limit theorem derives 
\begin{eqnarray}
    \frac{1}{T}\sum_{t=1}^T (\sqrt{T}p_{j,t})(\sqrt{T}\tilde{\xi}_t^{(i)})\sim N(0,\frac{1}{\sqrt{T}}). \nonumber
\end{eqnarray}
The numerical error $C_{\rm error}({\bX},\tilde{\bm\xi}^{(i)})$ is the sum of squared Gaussian random variables and follows the $\chi^2$ distribution with $r$ degrees of freedom \cite{dambre2012information}, 
\begin{eqnarray}
    C_{\rm error}({\bX},\tilde{\bm\xi}^{(i)})\sim \frac{1}{T}\chi^2(r). \nonumber
\end{eqnarray}

We choose the top $p$\% value $C_T$ of the distribution and determine the threshold to be the value multiplied by a scale $\sigma$, $C_{\rm th}=\sigma C_T$. 
We adopt $p=10^{-4}$ and $\sigma=2,3$ for the simulated QNRs and $p=10^{-2}$ and $\sigma=1$ for the QNRs implemented in real quantum machines. 
Using the threshold $C_{\rm th}$, we truncated the capacity $C$ as follows: 
\begin{eqnarray}
    C_{\rm truncate} = 
    \begin{cases}
        C & ({\rm if}~C\ge C_{\rm th}) \\
        0 & ({\rm otherwise})
    \end{cases}. 
\end{eqnarray}

\subsection*{Hardware Implementation}
Here we provide the details of the experiments on IBM superconducting quantum processors.
We prepared two types of input time-series with total length $T=200$; one in the symmetric range $u_{t}\in[-1,1]$ and the other in the asymmetric one $u_{t}\in[0,1]$.
We focused on these inputs to see the difference in the capacity depending on the input range as reported in \cite{kubota2021unifying}.

\begin{table*}
\caption{Device error parameters during the experiments and the total time-invariant capacities $\Ctottiv$ in our calculations. The medians of parameters are shown, where only the qubits constituting the QNR system are  considered.
} 
\label{tab:hardware_parameters}
\centering
\begin{tabular}{c|c|ccccc}
\hline
Input type & Parameter/Capacity& Kawasaki  & Toronto & Montreal & Manhattan\_p1 & Manhattan\_p2 \\
\hline\hline
 & CNOT error &0.0070 & 0.0083  &0.0095  &0.0259 &0.0161 \\
Symmetric&Readout error & 0.0095 & 0.0300 & 0.0140  &0.1499  &0.0183\\
&$\Ctottiv$& 0.3352 & 0.8388 & 1.035  & 0.3555 & 0.1973\\
\hline\hline
 & CNOT error &0.0070  & 0.0083 & 0.0097 & 0.0252 & 0.0163\\
Asymmetric&Readout error & 0.0095 & 0.0300 & 0.0138 & 0.1499 & 0.0183 \\
&$\Ctottiv$& 0.1811 & 0.6725 & 0.5662  & 0.2416 & 0.2099\\
\hline
\end{tabular}
\end{table*}

The experiments were performed using one Hammingbird-type device, \textit{ibmq\_manhattan} (Manhattan) and three Falcon-type devices, \textit{ibm\_kawasaki} (Kawasaki), \textit{ibmq\_montreal} (Montreal), and \textit{ibmq\_toronto} (Toronto).
Here, the QNR was built from 12 qubits in each device, where independent 2-qubit subsystems construct the whole system.
We considered two configurations, Manhattan\_p1 and Manhattan\_p2, to compare the capability of QNRs in different positions inside a large quantum Hammingbird-type device. 

To obtain the reservoir states, the systems were evolved by input-dependent unitary with hardware noise in Eq.~\eqref{eq:unitary}.
We performed the measurement in the Pauli $Z$ basis on each qubit, where we executed 8,192 shots for each time step.
Because the quantum state is affected by the projective measurement, we applied the unitary operators from the beginning to obtain the expectation values at every time step; that is, running $TS$ circuits in total, where $S$ is the number of measurement shots.
For each device and input type setting, we performed the same experiments 10 times. 
The device parameters used during the experiments are shown in Table~\ref{tab:hardware_parameters}.

\subsection*{Quantum Noise Simulation}
We describe three groups of quantum noises included in our study: (i) Pauli error, (ii) decoherence noise due to the coupling of physical qubits to the environment, and (iii) coherent or unitary noise arising from imperfect control or improper calibration.
For a specified error rate, we applied noises uniformly to all qubits after each input cycle. 

\textbf{Pauli error.}
The first common noise is the Pauli error, which can be used to simulate the bit-flip and phase-flip errors.
Assume that the error occurs with the probability $p$, then the bit-flip and phase-flip errors are defined though the Pauli X and Z operations, respectively, while the Pauli Y operation implies that both of the bit-flip and phase-flip errors occur.
We present here three representative Pauli errors: uniform depolarization, bit-flip, and phase-flip errors.

\textit{Uniform depolarization.---} A single-qubit uniform depolarization channel is described when all three types of Pauli errors have the same probability of occurring.
It can be described via the operator-sum representation
\begin{align}
    \rho \mapsto \cD(\rho) = \sum_{i=0}^3 K_i\rho K_i^{\dagger},
\end{align}
where $\{K_i\}$ are the Kraus operators defined by
\begin{align}
    K_0 = \sqrt{1 - p}I, &\quad K_1 = \sqrt{\frac{p}{3}}X, \nonumber\\
    K_2 = \sqrt{\frac{p}{3}}Y, &\quad K_3 = \sqrt{\frac{p}{3}}Z.
\end{align}

\textit{Bit-flip error.---}A single-qubit bit-flip channel is described by $\rho \mapsto \cD(\rho) = K_0\rho K_0^{\dagger} + K_1\rho K_1^{\dagger}$ with the following Kraus operators:
\begin{align}
    K_0 = \sqrt{1 - p}I,\quad K_1 = \sqrt{p}X.
\end{align}

\textit{Phase-flip error.---}A single-qubit phase-flip channel is described by $\rho \mapsto \cD(\rho) = K_0\rho K_0^{\dagger} + K_1\rho K_1^{\dagger}$ with the following Kraus operators:
\begin{align}
    K_0 = \sqrt{1 - p}I,\quad K_1 = \sqrt{p}Z.
\end{align}

\textbf{Decoherence noise.}
This error group simulates the interacting effects of the physical qubits with their surrounding environment, which are often described by non-unital operations. First, we consider the \textit{amplitude damping channel} associated with the thermal relaxation that occurs over time, which involves the exchange of energy between the qubits and their environment.
If we model the environment starting from the ground state $\ket{0}$, then this process drives the qubits towards the ground state $\ket{0}$.
Another noise process that is considered uniquely quantum mechanical is \textit{phase damping channel}, which describes the loss of quantum information without loss of energy. Here, the energy eigenstates are time-independent, but their phases are proportional to the eigenvalue. Therefore, the relative phases between energy eigenstates can be lost as evolving the system for a length of time~\cite{nielsen:2011:QCQI}.

Considering the damping rate $\gamma$, the damping channel can be described by $\rho \mapsto \cD(\rho) = K_0\rho K_0^{\dagger} + K_1\rho K_1^{\dagger}$, where the Kraus operators $K_0$ and $K_1$ are defined as follows for each channel.

\textit{Amplitude damping.---}A single-qubit amplitude damping channel is described by the following Kraus operators:
\begin{align}
K_0=\begin{pmatrix}
1 & 0 \\
0 & \sqrt{1 - \gamma}
\end{pmatrix}, \quad K_1 = \begin{pmatrix}
0 & \sqrt{\gamma}\\
0 & 0
\end{pmatrix}.
\end{align}

\textit{Phase damping.---}A single-qubit phase damping channel is described by the following Kraus operators:
\begin{align}
K_0=\begin{pmatrix}
1 & 0 \\
0 & \sqrt{1 - \gamma}
\end{pmatrix}, \quad K_1 = \begin{pmatrix}
0 & 0 \\
0 & \sqrt{\gamma}
\end{pmatrix}.
\end{align}

\textbf{Coherent or unitary noise.}
This error group describes the coherent errors, where the resulting operation is unitary but causes a modification from the target state. These errors can arise from imperfect control or improper calibration. We consider the following unitary noises to simulate the possible effect of coherent errors in our circuit model with
the noise rate $\epsilon = s \epsilon_{max}$, where $s \sim \mathrm{Uniform}(0,1)$ 
are sampled for each qubit.

\textit{Single-qubit over-rotation.---}A common case of coherent error describing single-qubit over-rotations:
\begin{align}
    \textup{RX}(\theta) \mapsto \textup{RX}(\theta ( 1+ \epsilon)),\quad \textup{RZ}(\theta) \mapsto \textup{RZ}(\theta ( 1+ \epsilon))
\end{align}
where the rotational angles are scaled by a factor of $1+\epsilon$. 

\textit{CNOT bias.---}To simulate over-rotation of CNOT gates, conditional $X$ rotation is scaled by a factor of $1+\epsilon$.
\begin{align}
    CNOT = 
    \begin{pmatrix}
    I & \\
    & \textup{RX}(\pi)
    \end{pmatrix} \mapsto  \begin{pmatrix}
    I & \\
    & \textup{RX}(\pi(1+\epsilon))
    \end{pmatrix} 
\end{align}

\textit{Unintended entangler (one-hop, two-hop)---}To simulate an unintended entangling between nearby qubits, the following small conditional rotation of factor $\epsilon$ is applied: 
\begin{equation}\label{eqn:unintentional_entangular}
 \begin{pmatrix}
    I & \\
    & \textup{RX}(\pi\epsilon)
    \end{pmatrix}.   
\end{equation}
We call them a one-hop entangler if these noises are applied between physically connected qubits, 
and a two-hop entangler if they are applied between physically non-connected qubits at a two-hop distance.

\subsection*{Echo State Property of QNR}
Echo State Property (ESP)~\cite{jaeger2002tutorial, manjunath2013echo, yildiz2012re} is a conventional prerequisite to guarantee the reproducibility of the computational task in RC.
However, ESP is not easily satisfied in the physical input-driven system due to the nature of the noisy environment and the short time scale of the system.
Here, we formulate the ESP and prove that the QNR constructed from the amplitude damping noise can satisfy the ESP.

Given an input sequence $u_0, u_1, u_2, \dots,$ the ESP of QNR can be formulated as
\begin{equation}\label{eqn:mes:esp}
   \|\hat{F}(\rho_0^{(1)}, \bu_T) - \hat{F}(\rho_0^{(2)}, \bu_T) \|_2 \rightarrow 0\quad \textup{as}\quad T \rightarrow \infty,
\end{equation}
for arbitrary initial states $\rho_0^{(1)}$ and $\rho_0^{(2)}$.
Here, $\bu_T = [u_0, u_1, u_2, \dots, u_T]$,  and $\hat{F}(\rho_0^{(1)}, \bu_T) = (\operatorname{tr}(Z_i\rho^{(1)}_T))_i \in \mathbb{R}^N$ and $\hat{F}(\rho_0^{(2)}, \bu_T) = (\operatorname{tr}(Z_i\rho^{(2)}_T))_i \in \mathbb{R}^N$ denote the reservoir states obtained after $T$ time steps when the quantum system starts from initial states $\rho_0^{(1)}$ and $\rho_0^{(2)}$, respectively. 
We consider a quantum version of ESP~\cite{chen:2019:dissipative,tran:2020:higherorder} under the trace distance of density matrices as
\begin{equation}\label{eqn:qesp}
    \|\rho^{(1)}_T- \rho^{(2)}_T \|_1 \rightarrow 0\quad \textup{as}\quad T \rightarrow \infty.
\end{equation}
If Eq.~\eqref{eqn:qesp} is satisfied, then we can also obtain Eq.~\eqref{eqn:mes:esp}.

First, we consider an amplitude damping channel with damping rate $\gamma$ ($\gamma < 1$) applying to the single-qubit case,
which can be represented as
\begin{align}\label{eqn:extbloch:amp}
    \begin{pmatrix}
       1\\
       r_x'\\
       r_y'\\
       r_z'
    \end{pmatrix} = \begin{pmatrix}
1 & 0 & 0 & 0 \\
0 & \sqrt{1-\gamma} & 0 & 0 \\
0 & 0 & \sqrt{1-\gamma} & 0\\
\gamma & 0 & 0 & 1-\gamma
\end{pmatrix}
    \begin{pmatrix}
       1\\
       r_x\\
       r_y\\
       r_z
    \end{pmatrix},
\end{align}
where $\br = (1, r_x, r_y, r_z)^\top$ denotes the extended Bloch vector form of a single-qubit state.
We denote the three-dimensional vector consisting of the last three elements in $\br^{(1)}_T- \br^{(2)}_T$ as $\Delta \br_T$, where $\br^{(1)}_T$ and $\br^{(2)}_T$ correspond to single-qubit states $\rho^{(1)}_T$ and $\rho^{(2)}_T$, respectively. We can verify that $\|\rho^{(1)}_T - \rho^{(2)}_T\|_1  = \|\Delta \br_T\|_2$ and $\Delta \br_{T}
 = \Gamma \Delta \br_{T-1}$, where
\begin{align}
\Gamma = \begin{pmatrix}
 \sqrt{1-\gamma} & 0 & 0 \\
  0 & \sqrt{1-\gamma} & 0 \\
  0 & 0 & 1-\gamma
\end{pmatrix}.
\end{align}
Therefore, $\|\Delta \br_{T}\|_2 \leq \| \Gamma \|_2 \|\Delta \br_{T-1}\|_2$,
where $ \| \Gamma \|_2$ denotes the spectral norm of $\Gamma$, which is 
 $\Gamma$'s largest singular value $\sigma_{\textup{max}}(\Gamma) = \sqrt{1-\gamma}$.

Next, we further consider the input-driven unitary transformation $\cU(u_t)$ that results in a rotation $\cR(u_T)$ in the non-extended Bloch representation before or after applying the amplitude damping channel.
In the matrix presentation, we can write $\Delta \br_{T}
 = \chi_T \Delta \br_{T-1}$, where $\chi_T$ includes both amplitude damping $\Gamma$ and rotation $\cR(u_t)$.
Given arbitrary square matrix $A$, the multiplication of any rotation matrix $R$ to either side of $A$ does not change the singular values of $A$.
Therefore, the spectral norm of $\chi_T$ is $\| \chi_T \|_2 = \|\Gamma\|_2 = \sqrt{1-\gamma}$ and
$\|\Delta \br_{T}\|_2 \leq \| \chi_T \|_2 \|\Delta \br_{T-1}\|_2 = \sqrt{1-\gamma} \|\Delta \br_{T-1}\|_2$.
Thus, 
\begin{align}\label{eqn:bloch:conver}
\|\Delta \br_{T}\|_2 \leq (1-\gamma)^{T/2} \|\Delta \br_{0}\|_2,
\end{align} 
which implies the ESP in Eq.~\eqref{eqn:qesp}.

Finally, we consider the $N$ qubit quantum states with amplitude damping channel applied to all qubits, and $\cU(u_t)$ is constructed from CNOT gates and single-qubit unitary channels as in Eqs.~\eqref{eq:unitary}\eqref{eq:unitary_2qubit}. 
Since the $i$-th element $z^{(1)}_{i,T} = \tr(Z_i\rho^{(1)}_T)$ of the reservoir states $\hat{F}(\rho^{(1)}_0, \bu_T)$ is obtained via performing the $Z$-axis measurement on the state $\rho^{(1)}_T$ of $i$-th qubit,
$z^{(1)}_{i,T}$ (and similarly with $z^{(2)}_{i,T}$) is equal to the $r_z$ of the extended Bloch vector for this $i$-th qubit.
From Eq.~\eqref{eqn:bloch:conver}, we can obtain $| \Delta z_{i,T} | = | z^{(1)}_{i,T} - z^{(2)}_{i,T}| \leq (1-\gamma)^{T/2} c_i$,
where $c_i \geq 0$ is a constant depending on initial states $\rho^{(1)}_0$ and $\rho^{(2)}_0$.
Therefore, 
\begin{align}\label{eqn:esp:rate}
    \|\hat{F}(\rho^{(1)}_0, \bu_T) -&\hat{F}(\rho^{(2)}_0, \bu_T) \|_2 \leq \sqrt{N} \max_i(| \Delta z_{i,T} |)\nonumber\\
    &\leq \sqrt{N(1-\gamma)^{T}}\max_i(c_i),
\end{align}
which implies the ESP defined in Eq.~\eqref{eqn:mes:esp}.
Furthermore, Eq.~\eqref{eqn:esp:rate} shows that a shorter term memory effect can be controlled via the damping rate $\gamma$.

\begin{acknowledgments}
This work is supported by MEXT Quantum Leap Flagship Program (MEXT Q-LEAP) Grant Nos. JPMXS0118067394 and JPMXS0120319794 and by JST CREST Grant No. JPMJCR2014, Japan.
\end{acknowledgments}

\section*{Author Contributions}
All authors conceived the research and contributed significantly to interpreting the results. 
T.K., Y.S., S.K. and Q.H.T conceived the model and prepared the manuscript.
T.K. performed the main analysis for the experimental data.
Y.S. developed the concept and designed the experiments in IBM quantum computers.
S.K. designed and implemented the simulation model.
K.N. and N.Y. supervised the research and contributed to the ideation and design of the research. 
All authors contributed to writing the manuscript.

\end{document}


\title{Supplemental Materials for ``Quantum Noise-Induced Reservoir Computing"}
\author{T. Kubota$^*$}
\author{Y. Suzuki$^*$}
\author{S. Kobayashi}
\author{Q. H. Tran}
\author{N. Yamamoto}
\author{K. Nakajima}%

\begin{abstract}
This supplementary material describes in detail the calculations, experiments introduced in the main text, and  additional figures.
The equation, figure, and table numbers in this section are
prefixed with S (e.g., Eq.\textbf{~}(S1) or Fig.~S1, Table~S1),
while numbers without the prefix (e.g., Eq.~(1) or Fig.~1, Table~1) refer to items in the main text.
\end{abstract}

\maketitle
\def\thefootnote{*}\footnotetext{These authors share first authorship.}\def\thefootnote{\arabic{footnote}}

\section{Details of IPCs in Benchmark Tasks and Simulated QNRs}

\begin{figure}
    \includegraphics[width=17cm]{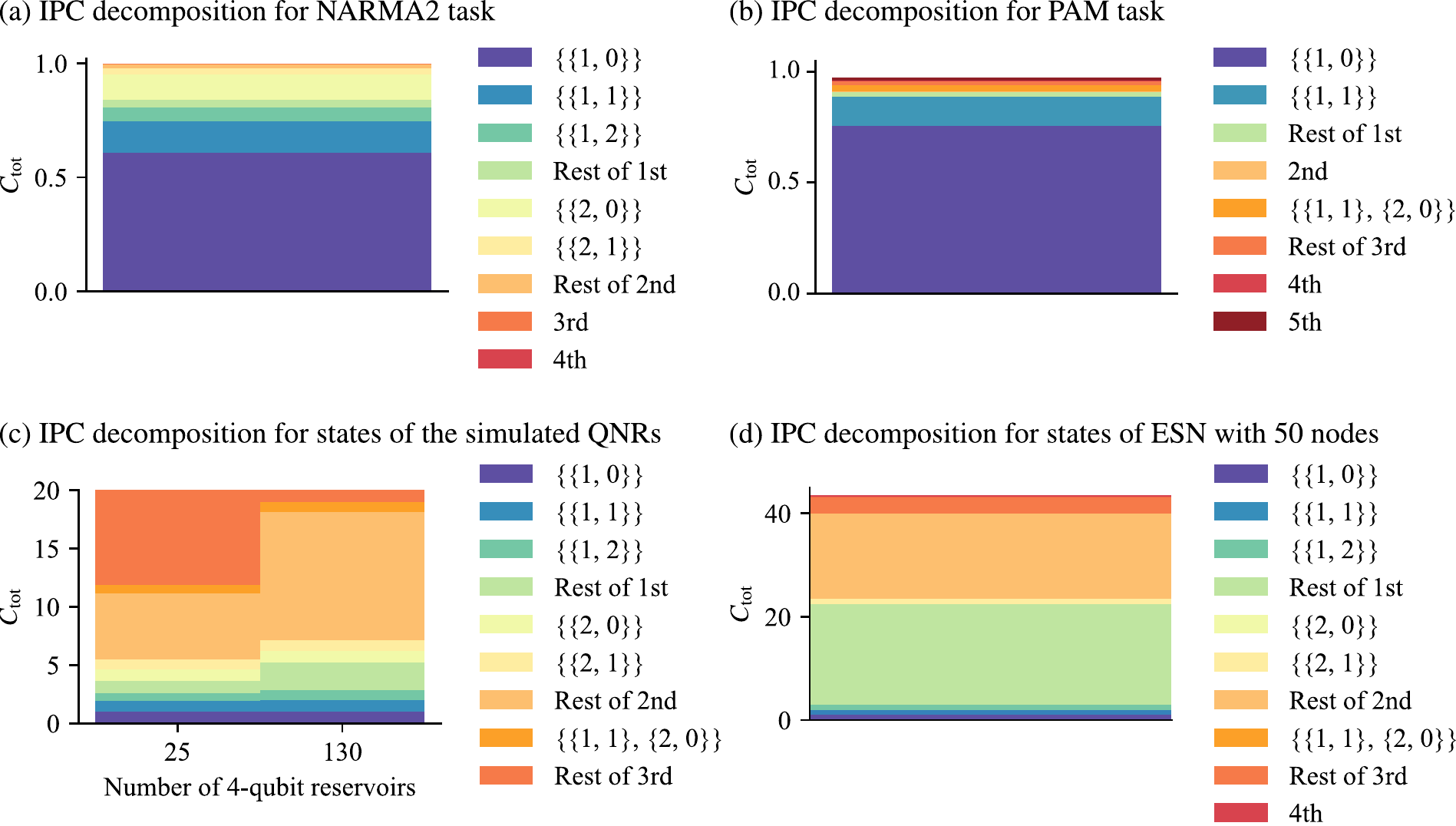}
    \caption{
        IPC decomposition for (a) NARMA2 task, (b) PAM task, (c) states of simulated QNRs in the main text, and (d) states of ESN with 50 nodes. The labels indicate representative combinations of $\{\{n_s, s\}\}$, where $n_s$ is the degree of polynomial, $s$ is the delayed time step of the input when the decomposed target is $\prod_s P_{n_s}(u_{n-s})$. The labels for other combinations whose corresponding IPC components were less than the threshold are omitted.
    }
    \label{fig:IPC_state_NARMA2}
\end{figure}

The time-invariant temporal information processing capacity (TIPC) is equivalent to information processing capacity (IPC) \cite{dambre2012information,kubota2021unifying}, which is a sufficient indicator to evaluate useful processed inputs for reservoir computing. 
For simplicity, 
we present here the details of calculated IPC components for the input\text{--}output relation in the NARMA2 task [Fig.~\ref{fig:IPC_state_NARMA2}(a)] and the PAM task [Fig.~\ref{fig:IPC_state_NARMA2}(b)]
 using the source code provided in Ref.~\cite{kubota2022github}. 
Note that we adopt uniform random input $u_t\in[-1,1]$ and Legendre polynomials for orthogonal bases. 
In addition, we remove numerical errors in IPC using 
the shuffle surrogate method~\cite{kubota2021unifying,kubota2022github}. 
This method prepares $N(=200)$ input time-series shuffled in the time direction and then computes $N$ surrogate IPCs $\{C_{{\rm shuffle},i}\}$ with the shuffled input to determine the threshold to $C_{\rm th}=\sigma\times\max_i{C_{{\rm shuffle},i}}$, where $\sigma(=1.2)$ is a scaling factor. Finally, we truncate the IPC as follows: 
\begin{eqnarray}
    C_{{\rm truncate},i} = 
    \begin{cases}
        C_i & ({\rm if}~C_i\ge C_{\rm th}) \\
        0 & ({\rm otherwise})
    \end{cases}. \nonumber
\end{eqnarray}

In the same manner as the TIPC decomposition $\Ctotdtiv$, the IPC decomposition is defined as the sum of IPCs for each order. 
We calculate the IPC decomposition for states of the simulated quantum noise-induced reservoirs (QNRs) when using 25 (130) instances of 4-qubit reservoirs in the spatial multiplexing setting [Fig.~\ref{fig:IPC_state_NARMA2}(c)] and for states of echo state network (ESN) with 50 nodes [Fig.~\ref{fig:IPC_state_NARMA2}(d)].
Here, the labels for IPC components indicate representative combinations of $\{\{n_s, s\}\}$, where $n_s$ is the degree of polynomial, $s$ is the delayed time step of the input with the decomposed target $\prod_s P_{n_s}(u_{n-s})$, where $P_n(u)$ represents the $n$th-order Legendre polynomial.

Our numerical calculation shows that the major components of IPC decomposition for the NARMA2 task is the first-order IPC  for $\{P_1(u_{t})\}, \{P_1(u_{t-1})\}$, and $\{P_1(u_{t-2})\}$ where the components are $0.585$, $0.145$, and $0.111$, respectively.
Those components are fulfilled in the IPC decomposition for states of the simulated QNRs and ESN, meaning that these reservoirs can solve the NARMA2 task.
For the PAM task, while the first-order IPCs still dominate, there are major IPC components for the combination $\{\{1,1\}, \{2,0\}\}\}$, which corresponds to the decomposed target $\{P_1(u_{t-1})P_2(u_{t})\}$.
We observe that these components appear in the IPC decomposition for states of the simulated QNRs but do not appear in the IPC decomposition for states of the ESN.
This observation confirms the better performance of the QNR compared with the ESN on the PAM task. Here, the QNR with a total of 520 computational nodes, meaning 130 instances of 4-qubit reservoirs, can solve the PAM task with the normalized root mean square $\textup{NRMSE}=0.21$, which slightly outperforms the ESN with less than 520 nodes ($\textup{NRMSE}>0.22$).

\section{Details of IPCs in QNRs implemented on IBM quantum machines}

\begin{figure}
    \includegraphics[width=17cm]{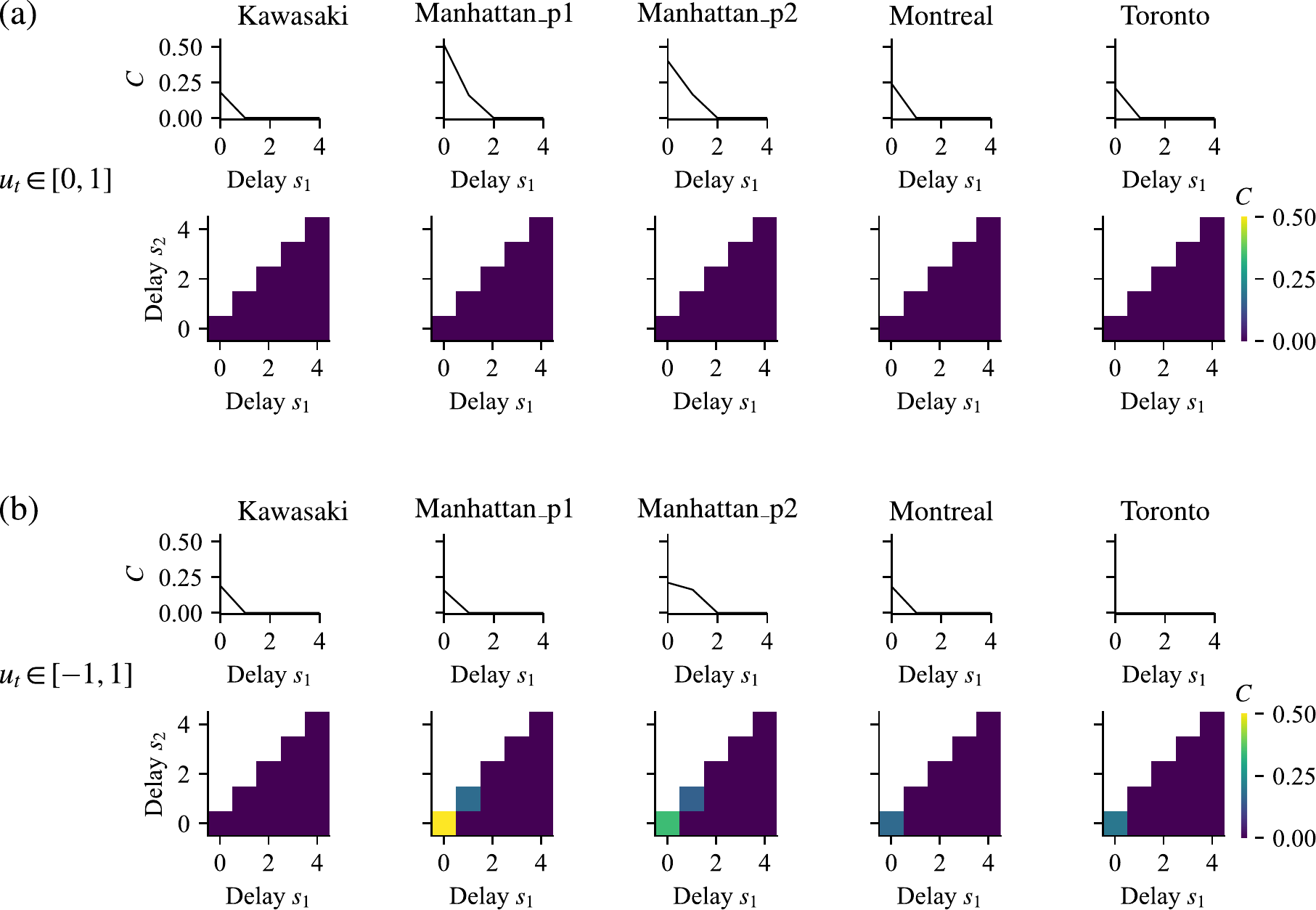}
    \caption{
        First-order capacities (first and third rows) and second-order capacities (second and fourth rows) of QNRs built on five IBM quantum machines for (a) asymmetric and (b) symmetric input. 
        The second-order capacities are zero in the asymmetric input case and present very short-term memories in the symmetric input case.
    }
    \label{fig:IPC_allmachines}
\end{figure}

\begin{figure}
    \includegraphics[width=17cm]{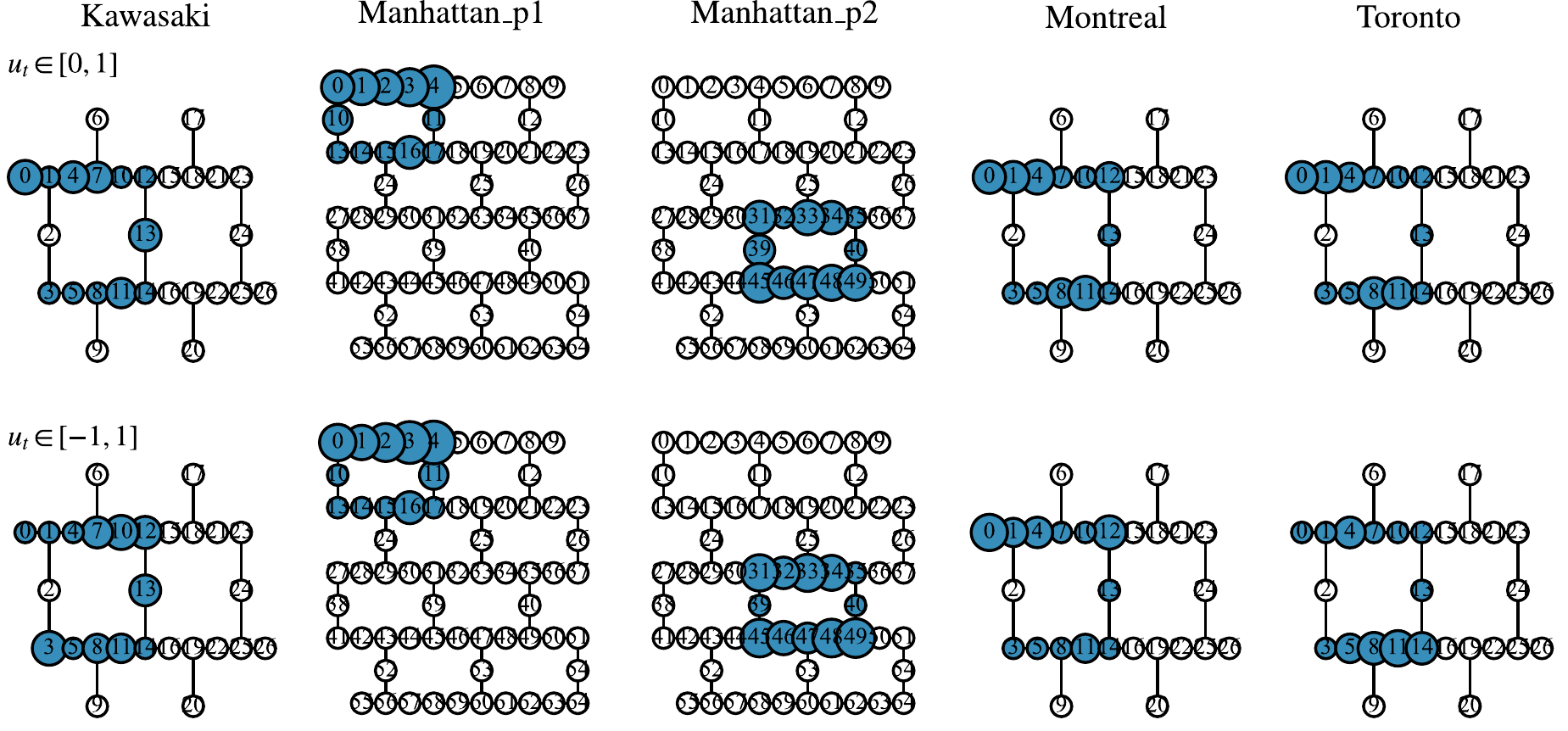}
    \caption{
    The total of first-order time-invariant capacities (represented by the size of nodes) contributed in each qubit in five IBM quantum machines for asymmetric (the first row) and symmetric (the second row) input. 
    }
    \label{fig:nodes_IPC_allmachines}
\end{figure}

We present here the details of calculated IPCs in our QNRs implemented on IBM quantum machines. 
Figure~\ref{fig:IPC_allmachines} depicts the first-order capacities (first and third rows) and second-order capacities (second and fourth rows) for asymmetric and symmetric input. 
Figure~\ref{fig:nodes_IPC_allmachines} shows the total of first-order time-invariant capacities (represented by the size of nodes) contributed with each qubit in five IBM quantum machines for asymmetric (the first row) and symmetric (the second row) input. 


\clearpage

\section{Echo state property (ESP) of QNR under the amplitude damping noise}
In the main text, we formulated the ESP of QNR model
given an input sequence $u_0, u_1, u_2, \dots,$ as
\begin{equation}\label{eqn:mes:esp}
   \|\hat{F}(\rho_0^{(1)}, \bu_T) - \hat{F}(\rho_0^{(2)}, \bu_T) \|_2 \rightarrow 0\quad \textup{as}\quad T \rightarrow \infty,
\end{equation}
for arbitrary initial states $\rho_0^{(1)}$ and $\rho_0^{(2)}$.
Here, $\bu_T = [u_0, u_1, u_2, \dots, u_T]$,  and $\hat{F}(\rho_0^{(1)}, \bu_T) = (\operatorname{tr}(Z_i\rho^{(1)}_T))_i \in \mathbb{R}^N$ and $\hat{F}(\rho_0^{(2)}, \bu_T) = (\operatorname{tr}(Z_i\rho^{(2)}_T))_i \in \mathbb{R}^N$ denote the reservoir states obtained after $T$ time steps when the quantum system starts from initial states $\rho_0^{(1)}$ and $\rho_0^{(2)}$, respectively. 

We have further proved that our QNR with $N$ qubits under the amplitude damping noise  applied to all qubits with damping rate $\gamma$ can satisfy the ESP via the following relation:
\begin{align}\label{eqn:esp:rate}
    \|\hat{F}(\rho^{(1)}_0, \bu_T) -&\hat{F}(\rho^{(2)}_0, \bu_T) \|_2 \leq c \sqrt{N(1-\gamma)^{T}},
\end{align}
where $c$ is a constant number depending on the initial states $\rho_0^{(1)}$ and $\rho_0^{(2)}$.

\begin{figure}
    \includegraphics[width=10cm]{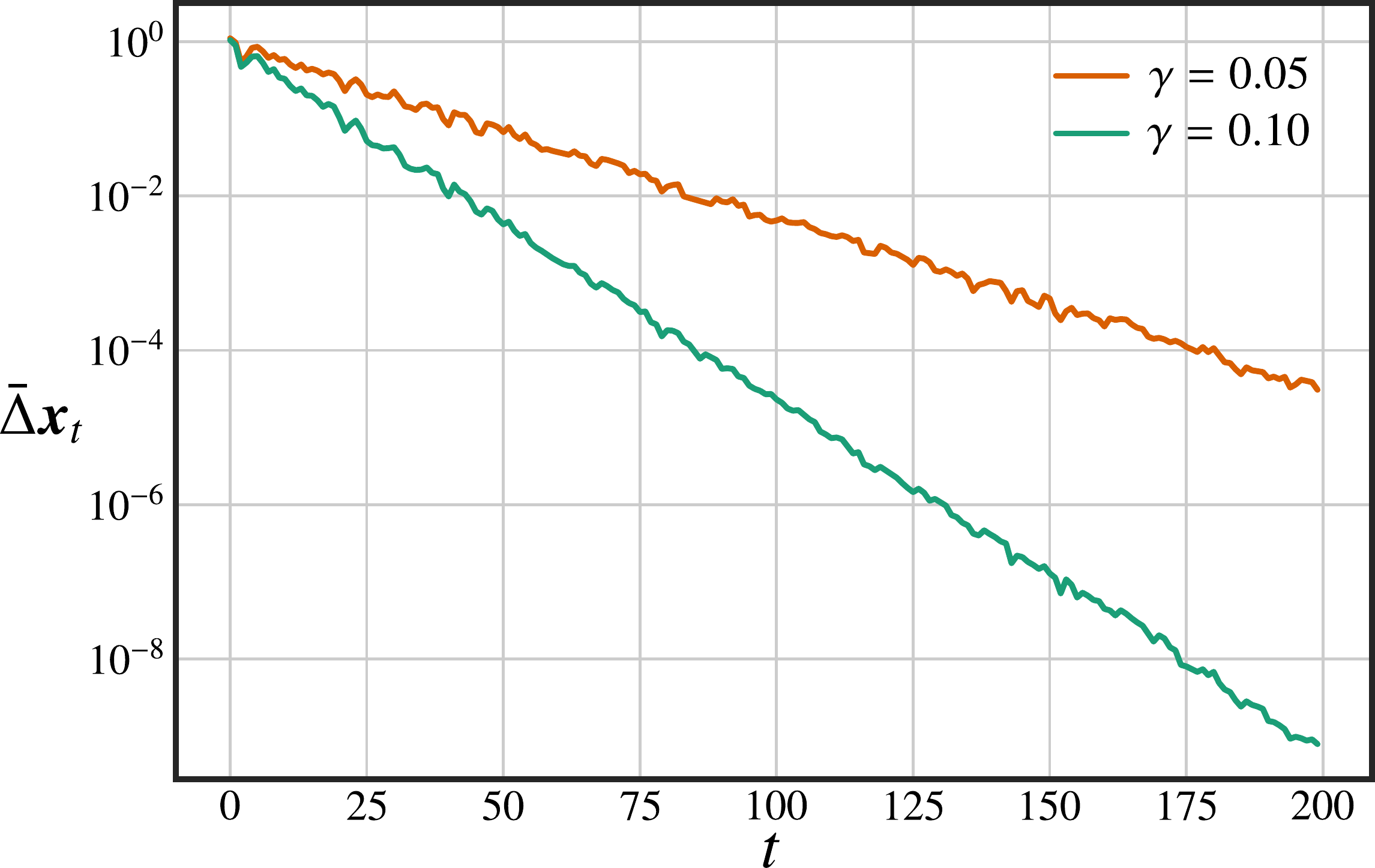}
    \caption{
    The average difference in reservoir states $\bar{\Delta} x_t$ of the QNR with amplitude damping rate $\gamma=0.10$ and $\gamma=0.05$. The y-axis is represented in the logarithmic scale, showing the value $\approx 10^{-4}$ for $\gamma=0.05$ and $\approx 10^{-8}$ for $\gamma=0.10$ around $T=175$. We can confirm that these results are consistent with Eq.~\eqref{eqn:esp:rate} because $\log(1 - 0.05)^{175} \approx 10^{-4}$ and $\log(1 - 0.1)^{175} \approx 10^{-8}$.
    }
    \label{fig:esp:decay}
\end{figure}

We conduct a numerical simulation to verify the relation in Eq.~\eqref{eqn:esp:rate}. 
We consider QNR with $N=4$ qubits and generate $M=20$ trials with the same random input sequence $u_0, u_1, \ldots,$ and amplitude damping noise, but different initial states $\rho^{(1)},\ldots,\rho^{(M)}$.
If the initial state of the quantum system is $\rho^{(m)}$, we obtain the corresponding time series reservoir states as  $\{\bx_0^{(m)},\bx_1^{(m)}, \cdots, \bx_T^{(m)}\}$,
where $\bx_t^{(m)} = \hat{F}(\rho_0^{(m)}, \bu_t) \in \mathbb{R}^N$. 
We compute the average difference between reservoir states for each time step with respect to the reservoir states generated from the initial state $\rho_0^{(1)}$ as 
\begin{align}
\bar{\Delta} \bx_t = \frac{1}{M-1}\sum_{m \neq 1} \|\bx_t^{(m)} - \bx_t^{(1)}\|_2.
\end{align}

Figure~\ref{fig:esp:decay} depicts the time variation of $\bar{\Delta} \bx_t$ with damping rates $\gamma=0.05, 0.10$. We confirm that $\bar{\Delta} x_t$ decays exponentially through time $t$ where the slopes of the log-plot are consistent with Eq.~\eqref{eqn:esp:rate}.


\providecommand{\noopsort}[1]{}\providecommand{\singleletter}[1]{#1}%